\documentclass[11pt]{elsart}
\usepackage{txfonts}
\usepackage{amssymb}
\usepackage[dvips]{graphicx}

\begin{document}
\begin{frontmatter}
\title{An effective local routing strategy on the BA network}
\author{Yu-Jian Li$^{a,b}$ }
\author{Zhen-Dong Xi$^{b,*}$ }
\author{Chuan-Yang Yin$^{a}$ }
\author{Bing-Hong Wang$^{a,*}$ }
\ead{bhwang@ustc.edu.cn}

\address
{$^{a}$ Department of Modern Physics and Nonlinear Science Center,
University of Science and Technology of China,  Hefei Anhui,
230026, PR China }

\address
{$^{b}$ Department of satellite measurement and control at sea of
China, Jiangyin Jiangsu,214400, PR China}

\begin{abstract}
In this paper, We propose a effective routing strategy on the basis
of the so-called nearest neighbor search strategy by introducing a
preferential delivering exponent $\alpha$. we assume that the
handling capacity of one vertex is proportional to its degree when
the degree is smaller than a cut-off value $K$, and is infinite
otherwise. It is found that by tuning the parameter $\alpha$, the
scale-free network capacity measured by the order parameter is
considerably enhanced compared to the normal nearest-neighbor
strategy. Traffic dynamics both near and far away from the critical
generating rate $R_c$ are discussed. We also investigate $R_c$ as
functions of $m$(connectivity density), $K$(cutoff value). Due to
the low cost of acquiring nearest-neighbor information and the
strongly improved network capacity, our strategy may be useful and
reasonable for the protocol designing of modern communication
networks.

\end{abstract}

\date{20080425}
\end{frontmatter}

\section{Introduction}
A variety of systems in nature can be described by complex
networks and the most important statistical features of complex
networks are the small-world effect and scale-free
property\cite{1,2,3}. It may serve as a very useful tool for
understanding nature and our society. Since the discovery of some
common interesting features of many real networks such as
small-world phenomena by Strogatz and Watts\cite{1} and Scale-free
phenomena by Albert and Barab\'{a}si \cite{2}, processes of
dynamics conducting on the network structure such as traffic
congestion of information now have drew more and more attention
from engineering and physical field
\cite{12,13,14,15,16,17,18,19,20,21,22}, due to the importance of
large communication networks such as WWW\cite{4} and Internet
\cite{5} in modern society. Evolution of networks structure and
interplay of traffic dynamics also play an important role in the
research of the traffic system including Internet, highway network
and so on , they want to understand and explain the process of
dynamics on the underlying structure.

Many previous excellent works focus on the evolution of structure
driven by the increment of traffic\cite{6,7,8} and some explore
how different topologies impact the traffic dynamics\cite{12,13}.
Some works\cite{10,11} gave several models to mimic the traffic
routing on complex networks by introducing randomly selected
source as well as particles (packets) generating rate and
destination of each particles\cite{18,19,20}. Those models define
the capacity of networks described by critical generating rate,
too. At this critical rate, a continuous phase transition from
free flow state to congested state occurs. In the free state, the
numbers of created and delivered particles are balanced, leading
to a steady state. While on the jammed state, the number of
accumulated particles increases with time due to the limited
delivering capacity or finite queue length of each vertex.

We believe that the study on the network search is very important
for traffic systems, for the existence of particles routing from
origin to destination and communication cost is very meaningful. a
few previous studies incorporate the search strategies and the
traffic processes on networks\cite{9,10,11,20}. In this paper, we
present a traffic model in which particles are routed only based
on local topological information with a single tunable parameter
$\alpha$. In order to maximize the nodes handling and delivering
capacity of the networks which can be measured by an introduced
order parameter $\etaup$, the optimal $\alpha$ is found out. We
also check the dynamic properties in the steady state for
different $\alpha$ including average number of particles versus
vertex¡äs degree, particles distribution and particles traveling
time distribution. The dynamics right after the critical
generating rate $R_c$ exhibits some interesting properties
independent of $\alpha$, which indicates that although the system
enters the jammed state, it possesses partial capacity for
forwarding particles. Our model can be considered as a
preferential walk among neighbor vertexes. We arrange the paper as
follows. In the following section we describe the model in detail,
in Sec.3 simulation results of traffic dynamics are provided in
both the steady and congested states, A conclusion and discussion
are given in the last section.

\section{Model}
our traffic model is described as follows: at each time step, there
are $R$ particles generated in the system, with randomly chosen
sources and destinations, and all vertexes can deliver at most $C$
particles toward their destinations, which is one of the most
interesting properties of the whole traffic network. The capacity of
each vertex is set to be proportional to its degree when the degree
is smaller than a cut-off value $K$, and be infinite otherwise, We
choose $C_i = k_i$. As a remark, there is difference between the
capacity of network and vertexes. The capacity of the whole network
is measured by the critical generating rate $R_c$ at which a
continuous phase transition will occur from free state to
congestion. The free state refers to the balance between created
particles and removed particles at the same time. When the system
enters the jam state, it means particles continuously accumulate in
the system and At last few particles can reach their destinations.
In order to describe the critical point accurately, we use the order
parameter\cite{9,10,11,20}:
\begin{equation}
\eta(R)=\lim\limits_{t\to \infty}\frac{C_i}{R}\frac{<\Delta
N_p>}{\Delta t}
\end{equation}

$\Delta N_p = N(t+\Delta t)-N(t)$ with $<\cdots>$ indicates average
over time windows of width $ \Delta t$ and $N_p(t)$ represents the
number of data particles within the networks at time t, here. For $R
< R_c$, $<\Delta N>= 0$ and $\etaup= 0$, indicating that the system
is in the free state with no traffic congestion. Otherwise for $R >
Rc$, $\etaup \to r$, where $r$ is a constant larger than zero, the
system will collapse at last. To navigate particles, each vertex
performs a local search among its neighbors. If a particle's
destination is found in the searched area, it will be delivered
directly to its target, otherwise, it will be forwarded to a
neighbor $j$ of vertex $i$ according to the probability:
\begin{equation}
\Pi_{i\to j}=\frac{k_j^\alpha}{\sum_lk_l^\alpha}
\end{equation}

here, the sum runs over the neighbors of vertex $i$ on the searched
area and $\alpha$ is an adjustable parameter studied by us in the
next context. Once a particle reaches its destination, it will be
canceled from the system.  As shown in Fig.1, the order parameter
versus generating rate $R$ by choosing different value of parameter
$\alpha$ is displayed. It is easy to Find that the capacity of the
system is not alike for different $\alpha$, thus, a natural question
is addressed: what is the optimal value of $\alpha$ for making the
network's capacity maximal in our model?

Many studies\cite{1,2,3} indicate that many communication networks
such as $Internet$ are not homogeneous like random or regular
networks. Barab\'{a}si and Albert proposed a famous model ($BA$ for
short) called scale-free networks\cite{2}, of which the degree
distribution is in good accordance with modern communication
networks, which has a power law distribution $P(k)\propto
k^{-\gamma}$. Our study is based on the so-called BA network, we
construct the network structure following the same method used in
Ref.\cite{2}: starting from $m$ fully connected vertexes, a new
vertex with $m$ links is added to the existing graph at each time
step according to the rule of preferential attachment i.e. the
probability of being connected to an existing vertex is proportional
to the degree of that vertex. Here, we choose $m = 5$ and network
size $N = 1000$ fixed for simulations.

We should also note that the queue length of each vertex is assumed
to be unlimited and the $FIFO$ (First in First out) discipline is
applied at each queue\cite{9,10,20,23}. Another important rule
called path iteration avoidance ($PIA$) is that a link between any
pair of vertexes is not allowed to be visited more than twice by the
same particle\cite{9,10,11}. Without this rule, the capacity of the
network is quite low due to many times' unnecessary visiting to the
same links by the same particles, which does not exist in the real
traffic systems. We note that this $PIA$ routing algorithm does not
damage the advantage of local routing strategy. If each particle
records the links it has visited, the $PIA$ can be easily performed.
One can find that this rule does not need the global topological
information. Therefore, we think this rule is rational and can
considerably improve the network capacity.

With the development of science and technology, the handling
capacity of the main central node can be set up large enough
artificially, that is why we propose the cutoff value $K$. We hold
that it is more practical and more reasonable.
\section{Simulations}
\begin{figure}
\begin{center}
\scalebox{1.2}[1.2]{\includegraphics{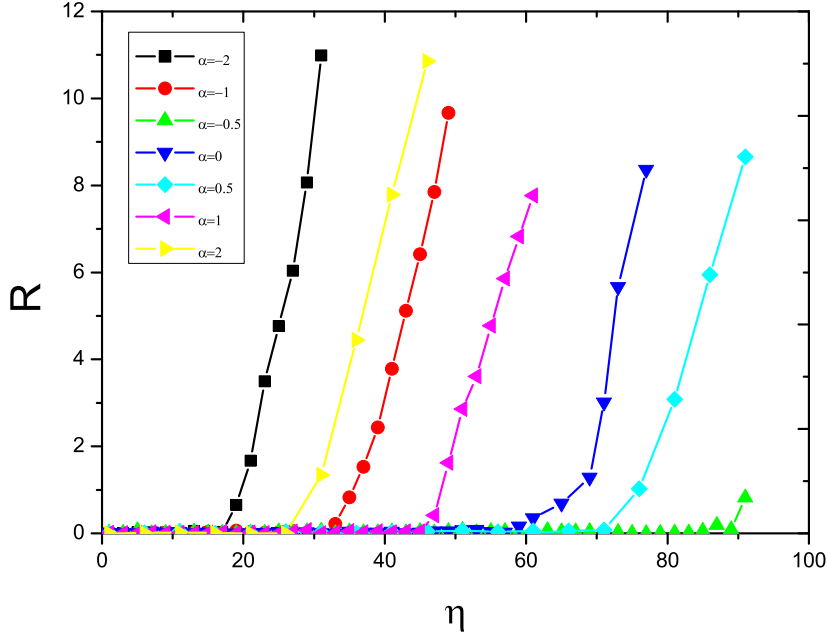}} \caption{(color
online). The order parameter $\etaup$ versus $R$ for BA network
with different free parameter $\alpha$. Other parameters are
networks size $N = 1000$, $K=1000$, and $m = 5$.}
\end{center}
\end{figure}
We have carried on the simulation under the definition of the model,
the order parameter $\etaup$ versus generating rate R with choosing
different value of parameter $\alpha$ is reported, As shown in Fig.
1. One can see that, for all different $\alpha$, $\etaup$ is
approximately zero when $R$ is small; it suddenly increases when $R$
is larger than the critical point $R_c$. It is easy to find that the
capacity of the system is not the same for different $\alpha$.For
the same $\etaup$, when $\alpha=-0.5$, the $R$ reaches its max. We
can preliminarily determine that it is the best situation.

\begin{figure}
\begin{center}
\scalebox{1.2}[1.2]{\includegraphics{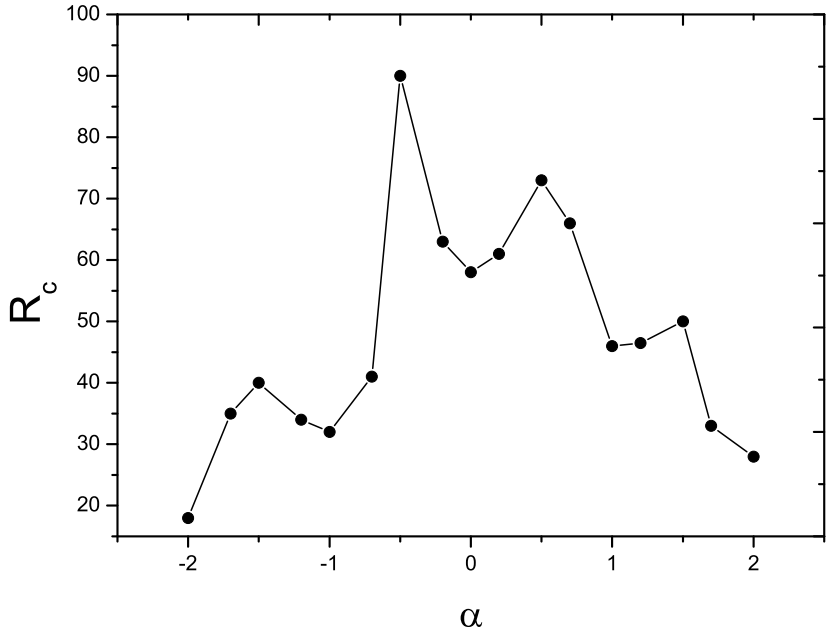}} \caption{The
critical $R_c$ versus $\alpha$ with network size $N =
1000$,$K=1000$. The maximum of $R_c$ corresponds to $\alpha=-0.5$
marked by a black solid line.}
\end{center}
\end{figure}
We also observed the handling capacity $R_c$ for different $\alpha$
in one system, one can read from Fig.2 that the tolerance is the
best when $\alpha=-0.5$. This is another strong evidence to show that
$\alpha=-0.5$ is a perfect point.One can read an multi-peak effect
at $\alpha=-1.5$, $-0.5$, $0.5$ and $1.5$. This is an interesting
phenomena,which we also gets in Fig.5. The perfect point is
$\alpha=0.5$ in that case, however, The nature of symmetry about
network can be reflected.

\begin{figure}
\begin{center}
\scalebox{1.2}[1.2]{\includegraphics{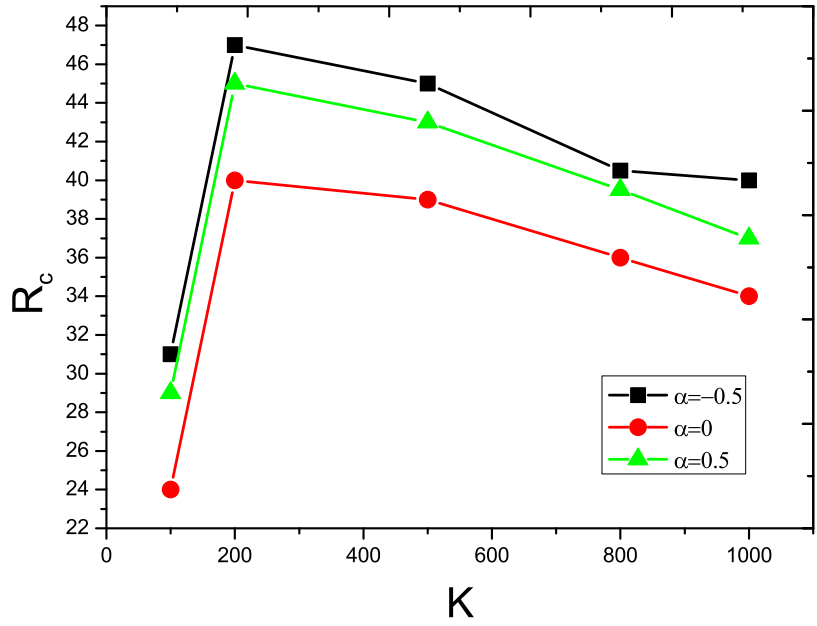}} \caption{(color
online). The variance of $R_c$ with the increasing of $K$}
\end{center}
\end{figure}

We have studied the critical point $R_c$ affected by the cutoff
value $K$, As shown in Fig.3. When the $K$ increases , the capacity
of BA network measured by $R_c$ considerably has the optimal
performance at $K=200$. Although $K$ is a variable parameter, the
system always reaches its best case at $\alpha=-0.5$.
\begin{figure}
\begin{center}
\scalebox{1.2}[1.2]{\includegraphics{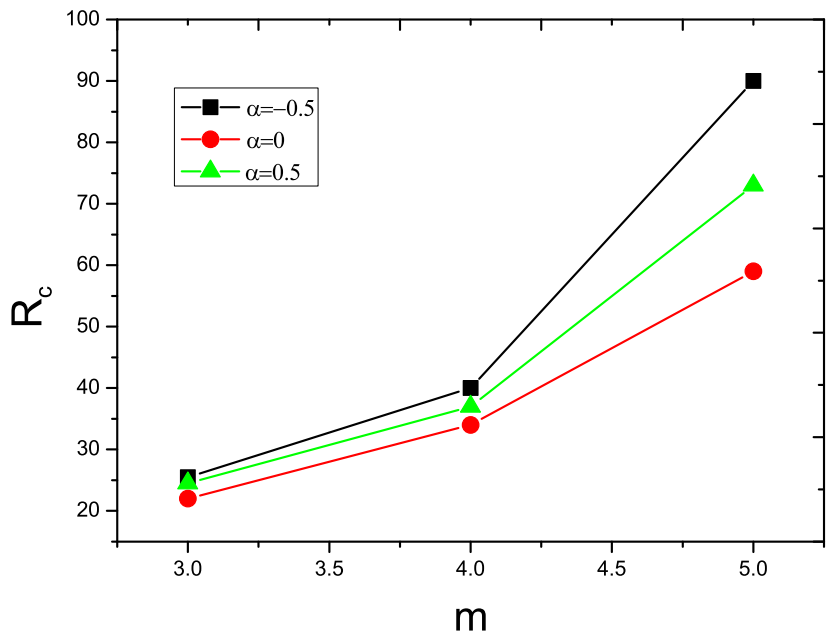}} \caption{(color
online). The variance of $R_c$ with the increasing of $m$}
\end{center}
\end{figure}
Furthermore, we study the critical point $R_c$ affected by the link
density of BA network. As shown in Fig.4, increment of $m$
considerably enhances the capacity of BA network measured by $R_c$
due to the fact that with high link density, particles can more
easily find their target vertexes. One can read the Fig.3 and Fig.4,
then know that the critical $R_c$ reaches its max when $\alpha
=-0.5$ at the same $K$ or $m$.

\begin{figure}
\begin{center}
\scalebox{1.2}[1.2]{\includegraphics{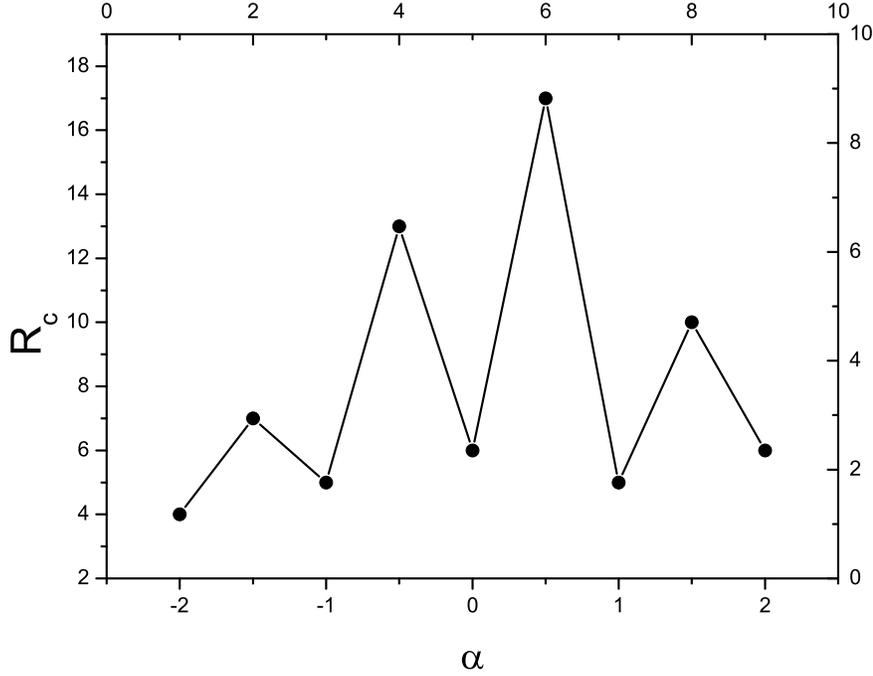}} \caption{The
critical $R_c$ versus $\alpha$ for $C_i=C\times k^2$ and $C=1$.
here, the network size $N=5000$}
\end{center}
\end{figure}

As shown in Fig.5 above, we have simulated the case when
$C_i=C\times k^2$ and we find that the system has the optimal
performance at $\alpha=0.5$. This simulation is based on the
practice that the capacity of the network is stronger with the
development of technology.

Simulation results demonstrate that the optimal performance of the
system corresponds to $\alpha =-0.5$. Compared to previous work by
Kim $et al$. \cite{18}, one of the best strategies is $PRF$
(preferential choice, which means the vertex with the larger degree
has the higher probability to receive particles) corresponding to
our strategy with $\alpha =-0.5$. By adopting this strategy a
particle can reach its target vertex most rapidly without
considering the capacity of the network. This result may be very
useful for search engine such as $google$, but for traffic systems
the factor of traffic jam cannot be ignored. Actually, average time
for the particles spending on the network can be also reflected by
the system capacity. It will indeed reduce the network's capacity if
particles spend too much time before arriving at their destinations.

\section{Conclusion}
We have introduced a new routing strategy only based on local
information, trying to give a solution to the problem of traffic
congestion in modern communication networks. Influenced by two
factors of each node's capacity and the cutoff value $K$, the
optimal parameter $\alpha=-0.5$ is obtained with maximizing the
whole system's capacity. In addition, the property that scale-free
network with occurrence of congestion still possesses partial
delivering ability suggests that only improving processing ability
of the minority of heavily congested vertexes can obviously enhance
the capacity of the system. The variance of critical value $R_c$
with the increment of $m$ and $K$ is also discussed. Our study may
be useful for designing communication protocols for large scale-free
communication networks due to the local information the strategy
only based on and the simplicity for application. The results of
current work may also shed some light on alleviating the congestion
of modern technological networks. Further work could be carried out,
for the queue length of each vertex is infinite.

\subsection*{Acknowledgement}
This work is funded by the National Basic Research Program of China
(973 Program No.2006CB705500), by the National Natural Science
Foundation of China(Grant Nos.60744003,10635040,10532060,10472116),
and by the Specialized Research Fund for the Doctoral Program of
Higher Education of China.

\end{document}